\documentclass[twocolumn]{aastex631}

\usepackage{soul}
\usepackage{hyperref}
\usepackage{CJK}
\usepackage{makecell}

\shorttitle{AASTeX v6.3.1 Sample article}
\shortauthors{Jiang et al.}
\graphicspath{{./}{figures/}}

\begin{document}
\begin{CJK*}{UTF8}{gbsn}
\title{Estimating the Mass Escaping Rates of Radius-valley-spanning Planets in the TOI-431 System via X-Ray and Ultraviolet Evaporation}

\author[0000-0002-6651-7809]{Xiaoming Jiang(蒋效铭)}\thanks{xmjiang@whu.edu.cn}
\affiliation{Department of Astronomy, School of Physics and Technology, Wuhan University, Wuhan 430072, China}

\author[0000-0002-5929-8951]{Jonathan H. Jiang}\thanks{Jonathan.H.Jiang@jpl.nasa.gov}
\affiliation{Jet Propulsion Laboratory, California Institute of Technology, Pasadena, CA 91109, USA}

\author[0000-0002-9020-7309]{Remo Burn}
\affiliation{Max-Planck Institute for Astronomy, 69117 Heidelberg, Germany}

\author[0000-0002-3567-6743]{Zong-Hong Zhu}\thanks{zhuzh@whu.edu.cn}
\affiliation{Department of Astronomy, School of Physics and Technology, Wuhan University, Wuhan 430072, China}








\begin{abstract}
TOI-431 system has 3 close-in exoplanets, which gives an ideal lab to study gas escape. 
In this study, we measure the XUV luminosity for TOI-431 with XMM-Newton/EPIC-pn and OM data, then calculate the fluxes for the planets in the system. We find that, TOI-431 b's $\rm F_{XUV,b}=$$70286^{+12060}_{-2611}$$\rm \ erg\ cm^{-2}s^{-1}$ is 75 times of TOI-431 d $\rm F_{XUV,d}=$$935^{+160}_{-35}$$\rm \ erg\ cm^{-2}s^{-1}$. 
Adopting the energy limit method and hydrodynamic code $ATES$ with a set of He/H ratios, we obtain the mass-loss rates of $10^{10.51^{+0.07}_{-0.02}}$ g s$^{-1}$ for TOI-431 b, $10^{9.14^{+0.07}_{-0.02}}$ and $10^{9.84\sim 9.94}$ g s$^{-1}$ for TOI-431 d. We predict the $2.93\sim 7.91 \%$ H I Ly$\alpha$ and $0.19\sim 10.65\%$ He I triplet absorption depths for TOI-431 d, thus its gas escaping is detectable in principle.
For both TOI-431 b and d, we select similar planets from the New Generation Planetary Population Synthesis (NGPPS) data. Then show that considering the mass-loss rates, TOI-431 b should be a naked solid planet, and TOI-431 d will likely maintain its gas envelope until the host star dies.
According to the formation and evolution tracks, we find that TOI-431 b's potential birthplace (0.1-2 AU) should be inner than TOI-431 d (2-12 AU). 
Our results are consistent with the interpretation of the radius valley being caused by atmospheric escape. The intrinsic reason may be their birthplace, which will determine how close they can migrate to the host star, then lose mass and result in the Fulton gap.
\end{abstract}


\section{Introduction}
Since the first discovery of exoplanets decades ago, a large number of exoplanets have been accumulated, and some parameter valleys appear in both observed exoplanet population and numerical simulation \citep{2021A&A...656A..69E, 2021A&A...656A..70E}, such as Fulton Valley (also called radius valley at $R_{p} \approx 1.7\ R_{\earth}$ in the distribution of planets' radii from the Kepler space mission \citep{2017AJ....154..109F}, the sub-Neptune desert \citep{2016A&A...589A..75M}, dearth of close-in stellar companions in survey \citep{2022AJ....163..232C}. The potential reasons for radius valley are gas-poor forming and atmospheric escape \citep{2013ApJ...775..105O,2017ApJ...847...29O,2017AJ....154..109F}. In the gas-poor forming story, rocky planets form after disks dissipate, thus they will not host a gas envelope. In the atmospheric escape scenario, some planets form and evolve in extremely hot environments, especially the close-in planets around young stars, they suffer violent stellar radiation. Their gas envelopes may be evaporated, inflated, and even stripped completely thus leaving the rocky component only.

\begin{table*}[t]
	\centering
	\caption{Planet parameters in TOI-431 system, reported by \cite{2021MNRAS.507.2782O}}
	\label{tab:example_table}
	\begin{tabular}{lccc} 
		\hline
		Parameters & TOI-431 b & TOI-431 c & TOI-431 d\\
		\hline
		Semi-major axis $a$ (AU) & $0.0113^{+0.0002}_{-0.0003}$ & $0.052\pm 0.001$ & $0.098\pm 0.002$\\
		Radius $R_p\rm(R_{\earth})$ & $1.28\pm 0.04$ & - & $3.29\pm 0.09$\\
            Mass $M_p\rm(M_{\earth})$ & $3.07\pm 0.35$ & $2.83^{+0.41}_{-0.34}$(M $sin$\ i) & $9.90^{+1.53}_{-1.49}$\\
            Bulk density $\rho \rm(g\ cm^{-3})$ & $8.0\pm 1.0$ & - & $1.36\pm 0.25$\\
            Equilibrium temperature $T_{eq}\rm(K)$ & $1862\pm 42$ & $867\pm 20$ & $633\pm 14$\\
		\hline
	\end{tabular}
\end{table*}

In theory, several dynamic mechanisms have been proposed to describe exoplanets' atmospheric escape \citep{2019AREPS..47...67O}. The most simple one uses the ratio of the gravitational potential of the gas particle to its thermal energy. In the upper atmosphere, gas particles can be heated and accelerated by high-energy photons, such that the gas particle can escape from the gravitational binding \citep{1963GeoJ....7..490O}. 
In the meantime, the density is low in the top atmosphere. Thus, the particles' mean free path can be longer than the scale height. If particles exceed the escape velocity, they will escape, namely, Jeans escape \citep{jeans1925dynamical}.
The close-in planets, suffer from violent radiation such that their atmospheric scale heights tend toward infinity, and the gas density becomes constant at large distances. It can be seen as the gas escaping to the outside region continually, so it corresponds to a hydrodynamic escape \citep{1958ApJ...128..664P}.
From the energy conservation view, the particles' thermal energy is dominated by incident energy which can be balanced against gravity. Under this assumption, an upper limit of the mass-loss rate can be calculated. The maximum mass-loss rate is obtained from the corresponding gravitational potential between the particle's initial altitude and the Roche lobe equal to the incident XUV energy with an efficiency factor, called energy-limited method \citep{1981Icar...48..150W,2007A&A...472..329E}.

In the observation aspect, atmospheric escape has been already confirmed. The most popular method is to compare the observed HI Lyman-$\alpha$ radius with the Roche radius. For example, HD209458 b whose broadband transit depth of about 1.5\% has been observed a 15\% atomic H I Lyman-$\alpha$ absorption, the corresponding radius has exceeded the Roche limit. This indicates the gas is escaping from the planet  \citep{2003Natur.422..143V,2000ApJ...529L..45C}.
Similarly, the second exoplanet with detected escaping gas, HD189733 b has a Lyman-$\alpha$ transit depth is 5.05$\%$ but its optic band depth is just 2.4$\%$ \citep{2010A&A...514A..72L, 2008MNRAS.385..109P}. 
Third, the close-in planet GJ436 b has a Lyman-$\alpha$ transit depth of $56.3\pm 3.5\%$ with an optical depth of 0.69$\%$ \citep{2015Natur.522..459E}.
Not only H I Lyman-$\alpha$, HD209458 b has been found deeper transit in 1180-1710 nm, O I ($13\pm 4.5\%$), C II ($7.5\pm 3.5\%$), but average depth is 5$\%$ \citep{2004ApJ...604L..69V}. Also, HD189733 b's optical transit depth is about 2.5$\%$, but it has been detected a $6.8\pm 1.8\%$ absorption at O I triplet (130.4 nm) \citep{2013A&A...553A..52B}. In addition, helium is a probe of gas absorption at 1083nm, WASP-107 b is found a $4.9\pm 1.1\%$ absorption depth, and the depths at other wavelengths are around $2\%$ \citep{2018ApJ...855L..11O,2018Natur.557...68S}.

Planets in the same system will share the same stellar history, and their orbits determine their environmental difference directly. Thus the multi-planet system is an ideal lab to study planets' thermal evolution and interaction to test the theory, such as TRAPPIST-1 \citep{2020SSRv..216..100T}, K2-3 \citep{2022AJ....164..172D}, and K2-136\citep{2023MNRAS.522.4251F}.
TOI-431 is a triple close-in planets system with two transit planets and an RV planet. Interestingly, these planets' sizes cover the Fulton gap, giving an ideal lab to study the relation between mass escaping and planets' sizes.

In this paper, we summarize the TOI-431 system in Section \ref{system}.
Then conduct XMM-Newton X-ray data reduction, light curve and SED analysis, and EUV conversion in Section \ref{observation}. Then estimate atmosphere mass-loss rates in Section \ref{Atmosphere mass-loss rate}. Additionally, using NGPPS to help us study the formation and evolution history of planets in Section \ref{Evolution history from NGPPS}. Finally, predict the gas escaping detectability for TOI-431 d in Section \ref{Gas escaping detectability}.

\section{Multiplanet system TOI-431}\label{system}
The multiplanet system TOI-431 was discovered by the Transiting Exoplanet Survey Satellite (TESS \citep{2015JATIS...1a4003R}) (TOI-431 b, d) and High Accuracy Radial Velocity Planet Searcher (HARPS \citep{2002Msngr.110....9P}) RV data (TOI-431 c) in 2021 \citep{2021MNRAS.507.2782O,2013PASP..125.1031B,2018A&A...616A...1G,2018MNRAS.475.4476W,1994SPIE.2198..362V,2016SPIE.9908E..84R,1999Msngr..95....8K,2012SPIE.8446E..88B,2019PASP..131k5003A,2021MNRAS.502.3704A}. 
The host star TOI-431 is $32.61\pm 0.01$ pc far away from Earth \citep{2021AJ....161..147B}, and it has been studied by \cite{2021MNRAS.507.2782O} in detail. Their results show that TOI-431 is a K3-type star, with a radius of $0.731 \pm 0.022 R_{\sun}$, mass of $0.78 \pm 0.07 M_{\sun}$, effective temperature of $4580 \pm 75 K$, and stellar age of $1.9 \pm 0.3 Gyr$ (SED fitting method) or $5.1 \pm 0.6 Gyr$ (elemental abundances method).

TOI-431 hosts 3 confirmed exoplanets, namely TOI-431 b, c, and d. TOI-431 b is a super-Earth-like planet with 1.28${\pm}$0.04 ${R_{\earth}}$ radius and 3.07${\pm}$0.35 ${M_{\earth}}$ mass on a 0.49 d orbit and is expected to be undergoing escape. TOI-431 d is a sub-Neptune, its radius is 3.29 ${\pm}$0.09 ${R_{\earth}}$ and mass is ${9.90^{+1.53}_{-1.49}}$ ${M_{\earth}}$. The third RV exoplanet is TOI-431 c, its transit hasn't been detected yet, thus we only know its ${M}$sin $i$ of ${2.83^{+0.41}_{-0.34}}$ ${M_{\earth}}$ \citep{2021MNRAS.507.2782O}.

\section{XMM-Newton observation}\label{observation}
\begin{figure*}[t]
    \centering	\includegraphics[width=1\textwidth]{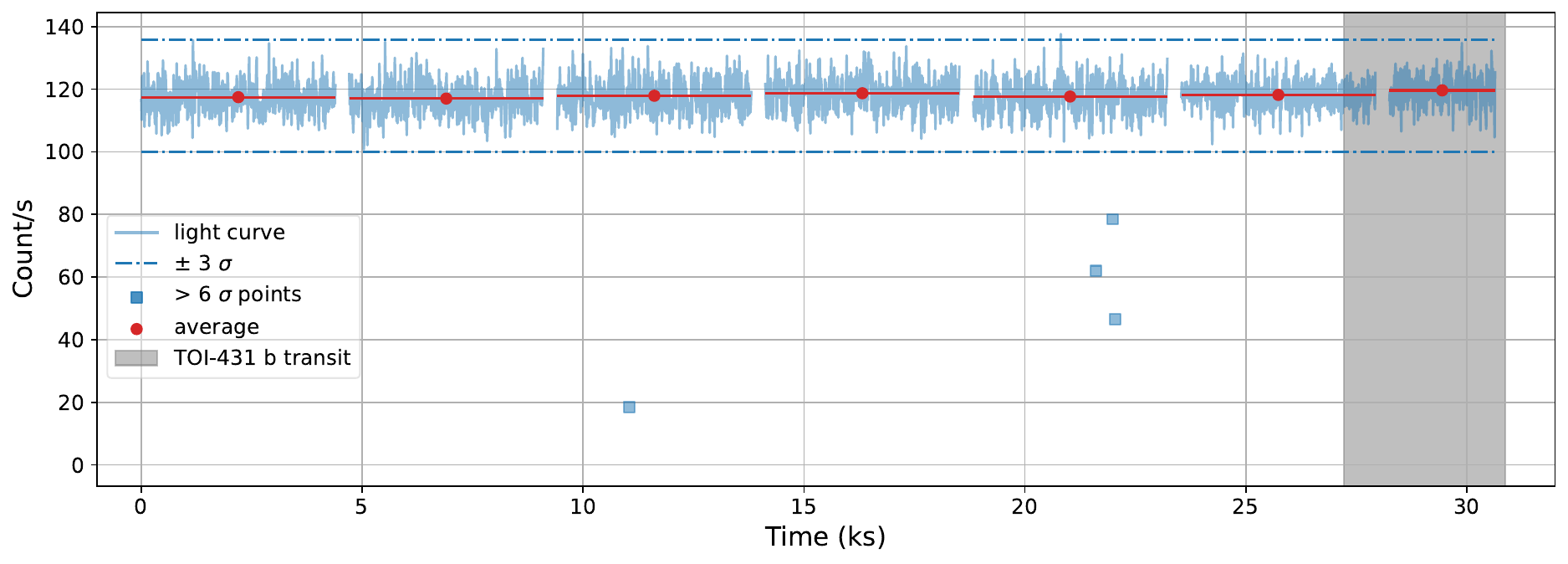}
    \caption{OM light curves of the TOI-431 (background subtracted). Grey shadow is the transit even of TOI-431 b.}
    \label{fig:light curve}
\end{figure*}
\subsection{Observation}
The host star TOI-431 was observed on 2021 August 28-29 by the XMM-Newton telescope with the EPIC (PN and MOS), RGS, and OM instruments  (Program PI: King, George, Observation ID: 0884680101), now the data is publicly available on the XMM-Newton Science Archive. To estimate the mass-loss rate, we need to verify the stellar luminosity stability and measure the incident XUV fluxes of planets. So we select the OM photometry data to plot the XUV light curve, and EPIC-pn spectrum data to estimate the XUV flux.
All the scientific exposures used in our study are summarised in Table \ref{Scientific exposures}.

\begin{table}[h]
\caption{Scientific exposures in this study}\label{Scientific exposures}

\begin{tabular*}{1\columnwidth}{lcccc} 
\hline
Exp.ID& Instrument& Mode& Filter& Exposure\\
\hline
S003 & EPN & Full Frame & THIN1 & 24519.21s\\
S006--12 & OM & FAST & UVW1 & 4400s$\times$6$+$2400s\\
\hline
\end{tabular*}
\end{table}

\subsection{SAS Data reduction}
After having the original EPIC-pn and OM data, we reduce it with the XMM-NEWTON Science Analysis Software (SAS) following the official step-by-step threads\footnote{https://www.cosmos.esa.int/web/xmm-Newton/sas-threads}. 

The main steps for the EPIC-pn spectrum include setting up the SAS environment, selecting source and background regions, and constructing the related response matrices. During determining when the light curve is low and steady, we notice that there are two outstanding peaks by the tail of the exposure, and we adjust the threshold to 0.5 counts s$^{-1}$ to define the "low background" intervals.
Due to the TOI-431's XUV flux being relatively weak, and to avoid lowering down spectrum resolution too much, when rebin the spectrum, we require at least 10 counts for each background-substracted spectral channel.

In the OM data reduction, to ensure the accuracy of the photometry, the tracking has been checked and shows there is no bad tracking frame. In each OM exposure, the tracking drift sizes of the majority of frames are lower than 0.4 pixels. And the integral drifts are rare to exceed 1 pixel. After the drift and standard flat correction, we sample the photon events of both the source and background with a time-bin size of 10 seconds.

\subsection{OM light curves}
To check the stability of the host star TOI-431, we drive its light curve as shown in Figure \ref{fig:light curve}. The mean count rate for the entire light curve is $117.46\pm 5.29$ counts s$^{-1}$, and the light curve clearly shows that during the XMM-Newton/OM observation, TOI-431's temporal variability within 3${\sigma}$ level, only 5 data points are out of the ${\pm}$3${\sigma}$ bias, and 4 of them lower than ${-6\sigma}$ boundary. The light curves also indicate no strong flare has been detected. Besides the stellar activity, during the whole observation, only TOI-431 b transits through the line of sight at the end of the OM light curve observation (transit mid-time from $NASA\ Exoplanet\ Archive$\footnote{https://exoplanetarchive.ipac.caltech.edu/index.html}: 2021/08/29 05:43), and doesn't affect the count rate significantly.

TOI-431's stellar activity is also studied by \cite{2021MNRAS.507.2782O}, they reported their monitoring of the stellar activity in WASP-South data (180+175+130 days in 2012-2014)\citep{2006PASP..118.1407P}, and found a $30.5\pm0.7$ d activity period with a 3 mmag amplitude of TOI-431, it may originate from starspot variation.

\subsection{XSPEC SED analysis}
\begin{figure}	 
    \includegraphics[width=1\columnwidth, angle=0]{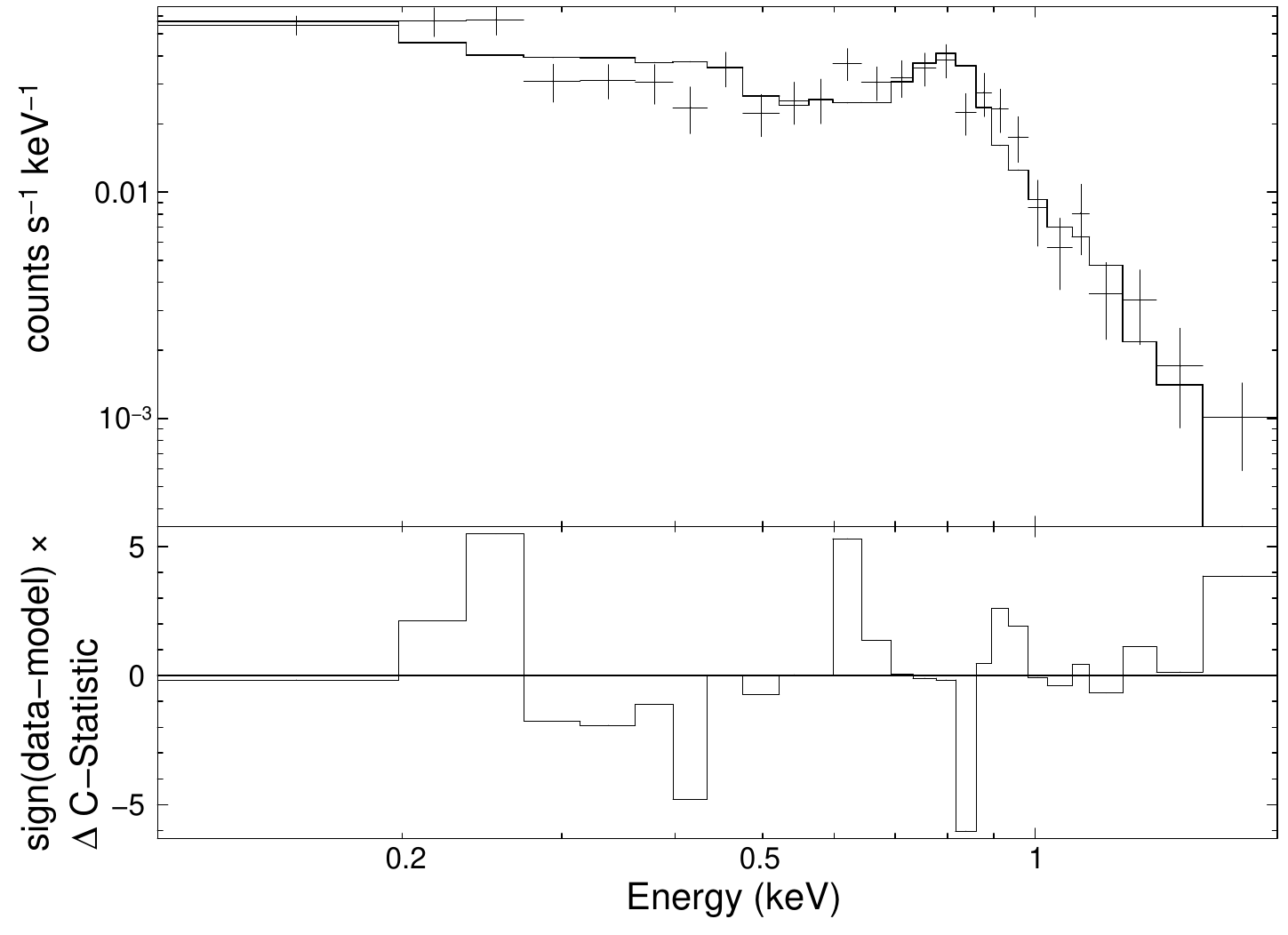}
    \caption{XMM-Newton EPIC-pn spectrum with the best-fit model and residual.}
    \label{fig:spectrum1}
\end{figure}

Once the EPIC-pn spectrum of TOI-431 has been reduced, we analyze it by the physical model using XSPEC with a similar method described in \cite{2019MNRAS.484L..49K}. Due to the low signal-to-noise ratio, we ignore the data over 2.5 keV.
The physical model has considered an optically thin thermal plasma model APEC \citep{2001ApJ...556L..91S} and inter-stellar median absorption model TBABS \citep{2000ApJ...542..914W}, with the initial column density assumption that the $\rm N_{HI}=1\times 10^{19}\ cm^{-2}$, which corresponds the local $\rm n_{HI}=0.1\ cm^{-3}$ \citep{2000ApJ...534..825R}, then free the column density during fitting. The metal abundance is frozen at 0.2, which is measured by \cite{2021MNRAS.507.2782O}. Then fit the spectrum with the C-statistic method, and estimate the 90\% confidence range for parameters. Finally, we obtain the unabsorbed X-ray flux at Earth $log10(\rm Flux_{\earth})=$$-13.30\pm0.03$ $\rm erg\ cm^{-2}\ s^{-1}$ in energy range of [0.124-12.4] keV, with a C-Statistic of 42.90 using 27 bins. The spectrum data, folded model, and contributions to C-Statistic are plotted in Figure \ref{fig:spectrum1}. 

However, we can see that there is at least one outstanding broad peak around 0.25 keV. To avoid the model approximating the peak by overestimation the total flux, we update the model in Section \ref{fit2}.

\subsection{XSPEC SED analysis II}\label{fit2}
\begin{figure}
	\includegraphics[width=1\columnwidth, angle=0]{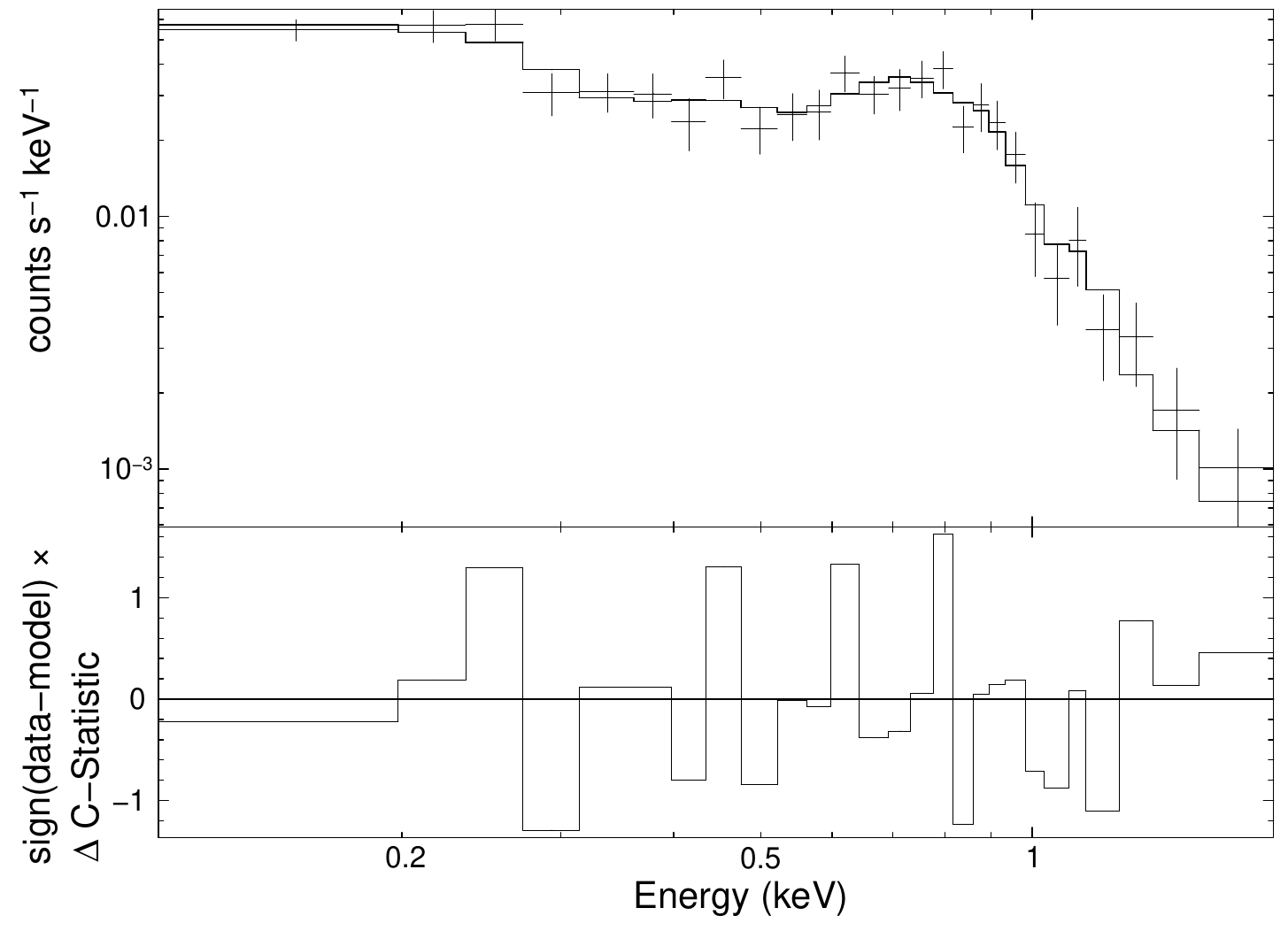}
    \caption{XMM-Newton EPIC-pn spectrum with the best-fit APEC, emission and absorption model with residual.}
    \label{fig:spectrum2}
\end{figure}
\begin{figure}
	\includegraphics[width=1\columnwidth, angle=0]{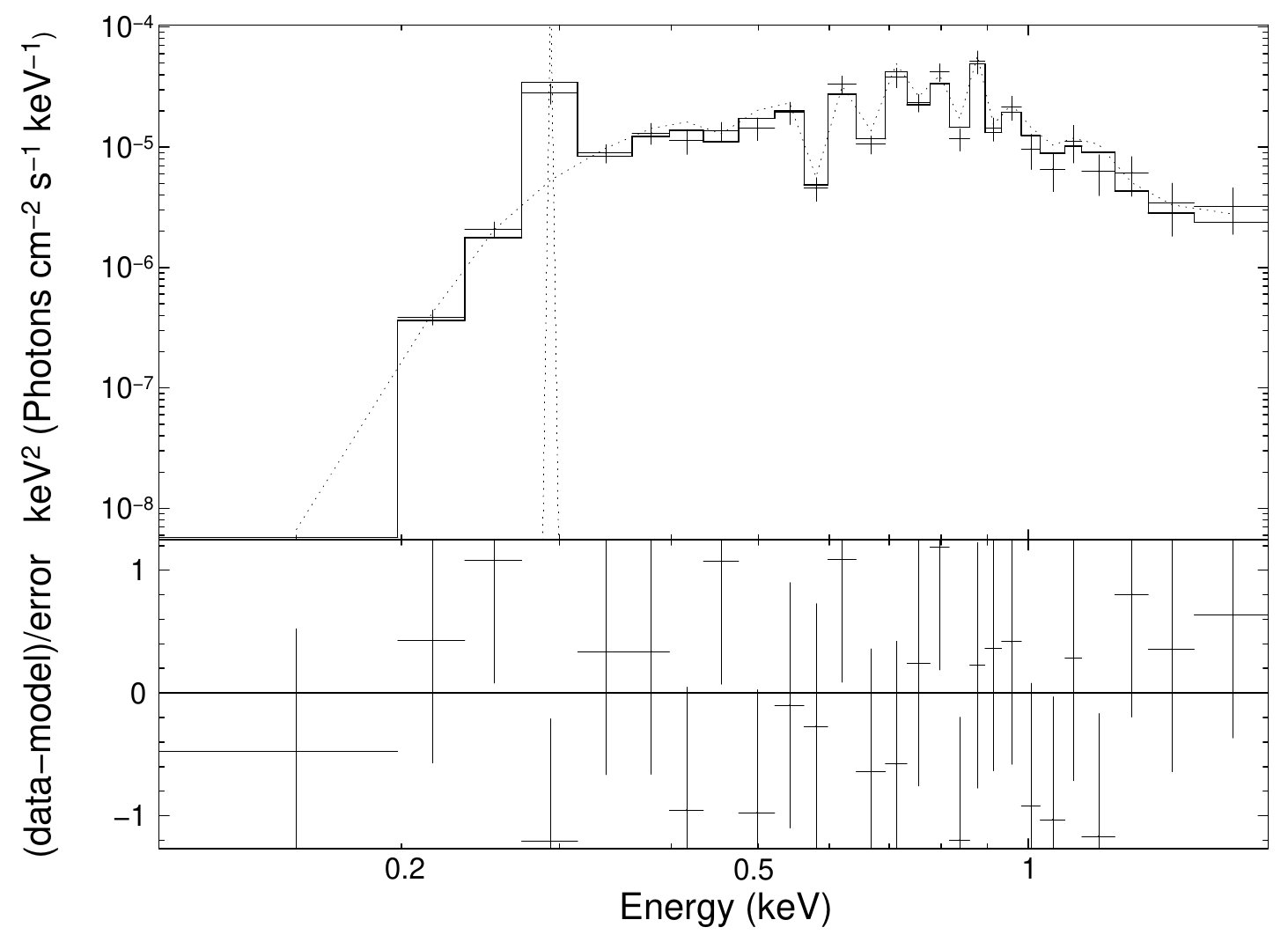}
    \caption{$\rm Energy^{2}$ unfolded plot for XMM-Newton EPIC-pn spectrum with the best-fit model.}
    \label{fig:spectrum3}
\end{figure}

We take a Gaussian emission and additional absorption components into the model for a more accurate and reasonable fitting. Initially, we specify the emission locations at 0.25 keV and free it in the fitting.
Then obtain a much better-fitting result as illustrated in Figure \ref{fig:spectrum2}, with a much lower C-Statistic of 15.73 using 27 bins. The result shows that the deviation of data and model is at the observation error level as shown in the lower panel of Figure \ref{fig:spectrum3}.
In the meantime, emission can match the data well at 0.28 keV, but the absorption is insignificant.

In addition, estimations of EUV flux in Section \ref{EUV flux} and hydrodynamic escaping in Section \ref{Hydrodynamic method} need different X-ray band fluxes. Thus we obtain the fluxes separately in energy ranges of [0.2-2.4] keV and [0.124-12.4] keV, with the same model parameters.
The corresponding X-ray fluxes with 90\% confidence ranges are listed in Table \ref{tab:XUV properties of TOI-431}. Because the spectrum is relatively X-ray soft, the high-energy region can not contribute much in flux, the X-ray fluxes in the two bands are comparable. And there is no data point above 2.5 keV, the [0.124-12.4] keV flux has a larger error range. 

Still, we can see a mall bump in 0.5-0.9 keV, \cite{2024MNRAS.530.3500K} linked it to Fe L shell emission, a characteristic of relatively young stars, $\geq$ 1 Gyr. Furthermore, in the following section \ref{System age issue}, our discussion based on NGPPS found that the age difference of 1.9 and 5.1 Gyr is not particularly significant for the planets, especially in the first 1 Gyr, when the planets have already reached a relatively stable state.

\subsection{EUV flux}\label{EUV flux}
The atmosphere's EUV scatter cross-section is greater than that in the X-ray band, thus EUV radiation is an important driver of atmospheric evaporation, it will affect the mass loss strongly \citep{2012MNRAS.425.2931O, 2015Icar..250..357C}. Unfortunately, EUV will be seriously absorbed by the interstellar medium, its observation is rare. The only way to reconstruct the EUV luminosity is using the luminosity relation between X-ray and EUV \citep{2015Icar..250..357C, 2018MNRAS.478.1193K}. For these reasons, we estimate the EUV luminosity and fluxes for the TOI-431 system with the following empirical relation from \cite{2018MNRAS.478.1193K}:
\begin{eqnarray}
     \rm \frac{F_{EUV}}{F_{X-ray}} &=&\rm \alpha (F_{X-ray})^\gamma
\end{eqnarray}
where $\rm \alpha = 1522\ erg\ cm^{-2}\ s^{-1}$ and $\gamma=-0.508$, for converting the [0.200-2.400] keV X-ray luminosity to [0.0136-0.124] keV EUV luminosity. Thus we obtain the X-ray and EUV luminosity as summarised in Table \ref{tab:XUV properties of TOI-431}, and these results are close to the values in \cite{2024MNRAS.530.3500K}.

\begin{table}[h]
	\centering
	\caption{Final model parameters, related emission measures, and XUV properties of TOI-431 system}
	\label{tab:XUV properties of TOI-431}
	\begin{tabular}{lc} 
            \hline
            Parameters & Value\\
  		\hline
		APEC kT (keV) & $0.346^{+0.162}_{-0.089}$\\
            APEC z & $0.162^{+0.061}_{-0.149}$\\
            APEC norm & $2.365^{+1.454}_{-4.007}\times10^{-4}$\\
            APEC EM ($\rm cm^{-3}$) & $3.139^{+4.962}_{-1.982}\times10^{51}$\\
            Gauss center (keV) & $0.280^{+0.125}_{-0.124}$\\
            Gauss sigma (keV) & $8.575\times10^{-6}\ ^{+0.081}_{-8.575\times10^{-6}}$\\
            Gauss norm & $1.014^{+9.252}_{-0.084}\times 10^{-4}$\\
            Gauss EM ($\rm cm^{-3}$) & $1.290^{+11.764}_{-0.107}\times10^{51}$\\
            TBabs ($\rm 10^{22}$) & $0.060^{+0.079}_{-0.046}$\\
            Gabs center (keV) & $0.611^{+0.054}_{-0.034}$\\
            Gabs sigma (keV) & $0.095^{+0.055}_{-0.031}$\\
            Gabs strength & $0.336^{+0.479}_{-0.202}$\\
 		\hline
		Band (keV) & $\rm log10(Flux_{\earth}) (erg\ cm^{-2}\ s^{-1}$)\\
		\hline
		X-ray [0.124-12.4] & $-13.33^{+0.16}_{-0.03}$\\
            X-ray [0.2-2.4] & $-13.33^{+0.07}_{-0.03}$\\
		\hline
		Band (keV) & $\rm log10(Luminosity) (erg\ s^{-1})$\\
		\hline
		X-ray [0.124-12.4] & $27.78^{+0.16}_{-0.03}$\\
            X-ray [0.2-2.4] & $27.78^{+0.07}_{-0.03}$\\
            EUV [0.0136-0.124] & $28.28^{+0.04}_{-0.01}$\\
		\hline
        Band (keV) & $\rm Flux_{TOI-431 b,d} (erg\ cm^{-2}\ s^{-1})$\\
		\hline
		X-ray [0.124-12.4] & $16653^{+7462}_{-989}$\ \ $221^{+99}_{-13}$\\
            X-ray [0.2-2.4] & $16657^{+3031}_{-1008}$\ \ $221^{+40}_{-13}$\\
            EUV [0.0136-0.124] & $53633^{+4598}_{-1622}$\ \ $713^{+61}_{-22}$\\
            XUV [0.0136-12.4] & $70286^{+12060}_{-2611}$\ \ $935^{+160}_{-35}$\\
		\hline
	\end{tabular}
\end{table}

\section{Atmosphere mass-loss rates}\label{Atmosphere mass-loss rate}
\begin{table}[h]
	\centering
	\caption{Mass-lost rates in TOI-431 system}
	\label{tab:mass-loss rate}
	\begin{tabular}{lcc} 
		\hline
		Parameters & TOI-431 b & TOI-431 d\\
		\hline
            \makecell*[l]{\hspace{-4em} Energy limit\\ \hspace{-4em} $\dot{M}$ log10(g/s)} & $10.25^{+0.07}_{-0.02}$ & $9.1^{+0.07}_{-0.02}$\\
            \\
            Roche lobe parameter K & 0.556 & 0.908\\
            \\
            \makecell*[l]{\hspace{-4em} Energy limit Roche\\ \hspace{-4em} $\dot{M}$ log10(g/s)} & $10.51^{+0.07}_{-0.02}$ & $9.14^{+0.07}_{-0.02}$\\
            \\
            \makecell*[l]{\hspace{-4em} $ATES$ $\dot{M}$ log10(g/s) \\ \hspace{-4em} (He/H=100)} & NaN & $9.85^{+0.08}_{-0.01}$\\
            
            \makecell*[l]{\hspace{-4em} $ATES$ $\dot{M}$ log10(g/s) \\ \hspace{-4em} (He/H=50)} & NaN & $9.84^{+0.08}_{-0.01}$\\
            
            \makecell*[l]{\hspace{-4em} $ATES$ $\dot{M}$ log10(g/s) \\ \hspace{-4em} (He/H=20)} & NaN & $9.84^{+0.08}_{-0.02}$\\
            
            \makecell*[l]{\hspace{-4em} $ATES$ $\dot{M}$ log10(g/s) \\ \hspace{-4em} (He/H=10)} & NaN & $9.84^{+0.08}_{-0.02}$\\
            
            \makecell*[l]{\hspace{-4em} $ATES$ $\dot{M}$ log10(g/s) \\ \hspace{-4em} (He/H=4)} & NaN & $9.84^{+0.06}_{-0.02}$\\
            
            \makecell*[l]{\hspace{-4em} $ATES$ $\dot{M}$ log10(g/s) \\ \hspace{-4em} (He/H=2)} & NaN & $9.84^{+0.06}_{-0.02}$\\
            
            \makecell*[l]{\hspace{-4em} $ATES$ $\dot{M}$ log10(g/s) \\ \hspace{-4em} (He/H=1)} & NaN & $9.84^{+0.08}_{-0.01}$\\
            
            \makecell*[l]{\hspace{-4em} $ATES$ $\dot{M}$ log10(g/s) \\ \hspace{-4em} (He/H=0.5)} & NaN & $9.85^{+0.07}_{-0.02}$\\
            
            \makecell*[l]{\hspace{-4em} $ATES$ $\dot{M}$ log10(g/s) \\ \hspace{-4em} (He/H=0.2)} & NaN & $9.87^{+0.06}_{-0.02}$\\
            
            \makecell*[l]{\hspace{-4em} $ATES$ $\dot{M}$ log10(g/s) \\ \hspace{-4em} (He/H=0.083)} & NaN & $9.90^{+0.06}_{-0.02}$\\
            
            \makecell*[l]{\hspace{-4em} $ATES$ $\dot{M}$ log10(g/s) \\ \hspace{-4em} (He/H=0.05)} & NaN & $9.91^{+0.06}_{-0.01}$\\
            
            \makecell*[l]{\hspace{-4em} $ATES$ $\dot{M}$ log10(g/s) \\ \hspace{-4em} (He/H=0.02)} & NaN & $9.93^{+0.05}_{-0.01}$\\
            
            \makecell*[l]{\hspace{-4em} $ATES$ $\dot{M}$ log10(g/s) \\ \hspace{-4em} (He/H=0.001)} & NaN & $9.94^{+0.06}_{-0.01}$\\
            
            \makecell*[l]{\hspace{-4em} \cite{2021MNRAS.507.2782O}'s\\ \hspace{-4em} $\dot{M}$ log10(g/s)} & 10-11 & 8.7-9.7\\

            \makecell*[l]{\hspace{-4em} \cite{2024MNRAS.530.3500K}'s\\ \hspace{-4em} $\dot{M}$ log10(g/s)} & NaN & 9.55-9.89\\
            
		\hline
	\end{tabular}
\end{table}

\subsection{Energy limit method}
The atmosphere mass-loss rates are usually estimated by the energy-limit method, and the incident X-ray flux is seen as the energy source of gas escaping \citep{2019MNRAS.484L..49K,2022A&A...661A..23F}. Then combining the correction of the Roche sphere, the estimation formula is obtained \citep{1981Icar...48..150W,2007A&A...472..329E}:

\begin{eqnarray}
    \dot{M}&=&\frac{\eta\beta^{2}\pi F_{XUV}R_P^{3}}{GKM_{P}}\\
    K&=&1-\frac{3}{2\xi}+\frac{1}{2\xi^3}\\
    \xi&=&\frac{r_{Rl}}{r_{pl}}\\
    r_{Rl}&\approx&(\frac{\delta}{3})^\frac{1}{3}d\\
    \delta &=& \frac{m_{pl}}{m_{star}}
\end{eqnarray}
In the formula, $\eta$ represents the energy conversion efficiency, and usually be assumed as 0.15, $\beta = 1.1$ is the ratio of $R_{optic}$ and $R_{XUV}$, it indicated in XUV band planet may have a larger radius due to inflated gas envelopes can absorb the XUV radiation. These assumptions are adopted from \cite{2021MNRAS.500.4560P,2022A&A...661A..23F}. Though \cite{2022A&A...661A..23F} neglect the Roche lobe, approximate the parameter $K$ to 1, we calculate $K$ carefully to consider the Roche lobe as described in \cite{2007A&A...472..329E}, which can enhance the escaping rate for the most close-in planets. The final results of the energy limit method are collected in Table \ref{tab:mass-loss rate}.

\subsection{Hydrodynamic method}\label{Hydrodynamic method}

However, the energy limit method has a large uncertainty, because the energy absorption efficiency $\eta$ and the XUV-optic radius ratio of the planet $\beta$ are unknown, and the planet's internal heat source may also contribute to the gas escaping. At this point, hydrodynamic escaping should be considered.

We calculate the current mass-loss rate with the photoionization hydrodynamics code $ATES$\footnote{https://github.com/AndreaCaldiroli/ATES-Code} \citep{2021A&A...655A..30C}. The $ATES$ is designed for numerical solving conservative, one-dimensional Euler equations and calculating the temperature, density, velocity, and ionization fraction profiles of highly irradiated planetary atmospheres, along with the current, steady-state mass loss rate. It considers a primordial atmosphere composed of atomic hydrogen and helium. Assume the He to H number density ratio of 0.083 as the default, and to compare the He/H ratio range in \cite{2020A&A...638A..61P,2020A&A...636A..13L, 2021A&A...647A.129L} we include a set of He/H ratios in addition, it will be discussion in Section \ref{He/H ratio}. It would be possible that there is also water in the envelope but we do not consider this yet.

We input the planetary mass, radius, equilibrium temperature, orbital distance, host star's mass, X-ray and EUV luminosity to $ATES$ for both TOI-431 b and d then run the code with the assumed He/H ratios, default total number density at planet radius of $10^{14} cm^{-3}$ and power-law spectral index of -1 which was validated by \cite{2021A&A...655A..30C}. Consequently, the TOI-431 d obtains a stable solution, but the TOI-431 b does not, its situation will be discussed in Section \ref{toi-431 b discussion}.

Overall, different He/H ratios give a mass-loss rate variance range of $25\%$ as summarized in Table \ref{tab:mass-loss rate}, and our results are also consistent with the rough estimations in \cite{2021MNRAS.507.2782O}, and recent work by \cite{2024MNRAS.530.3500K}, the He/H ratio issue will be discussed in detail in Section \ref{He/H ratio}. For a conservable mass-loss rate, the evolution history section \ref{Evolution history from NGPPS} will be based on the result of the He/H ratio of 0.083, which gives a median close value.

\section{Gas escaping detectability}\label{Gas escaping detectability}
\begin{figure*}[t]
\centering
	\includegraphics[width=1\textwidth]{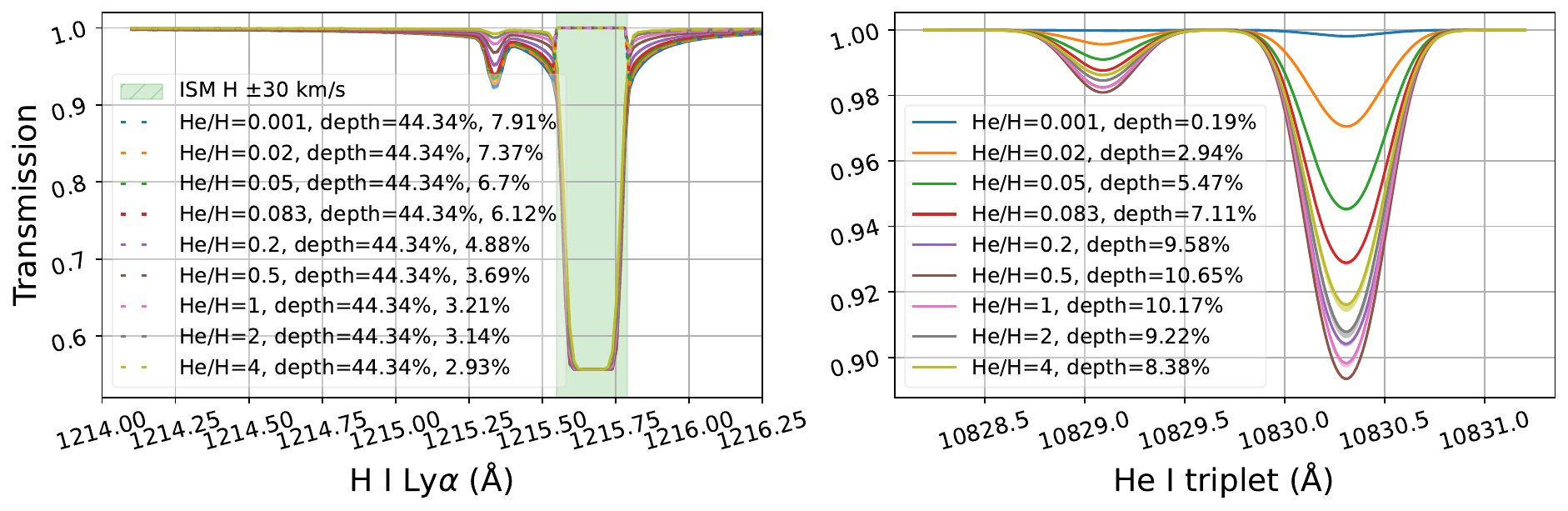}
    \caption{Theoretical intrinsic (solid) and ISM absorbed (dashed) transmission profiles of H I Ly$\alpha$ (the ISM absorbed depths values are listed after the intrinsic depths) and He I triplet for TOI-431 d with the uncertainties range derived from the mass-loss rate uncertainties. The theoretical intrinsic maximum H I Ly$\alpha$ transit depths are almost unchanged $44.34\%$ at 1215.6 $\rm \AA$, but with the ISM absorption, transit depths will decrease continually as He/H ratio goes up, and much weaker than the intrinsic transit depth. The He I triplet absorption depths at 10830.3 $\rm \AA$ will increase up to 10.65 $\%$ at He/H of 0.5, then decrease as He/H goes up.}
    \label{figures/transmission_all_ism_2}
\end{figure*}
In the above discussion, we estimate the mass-loss rates of the TOI-431 planets via indirect methods. Unlike other observed gas escape planets (e.g, HD209458 b \citep{2003Natur.422..143V, 2008A&A...483..933E, 2018ApJ...855L..11O}, GJ 436 b \citep{2015Natur.522..459E, 2018ApJ...855L..11O}, WASP-107 b \citep{2018Natur.557...68S}, HAT-P-32 b \citep{2023SciA....9F8736Z}), TOI-431 planets' gas escaping hasn't been directly observed yet. To predict the maximum absorption of the escaping gas and estimate whether it would be observable, we use the tool in $ATES$ to calculate both H I Ly$\alpha$ and He I triplet transmission for TOI-431 d. With the assumed impact parameter $b = 0$, and the same He/H assumptions and results from Section \ref{Hydrodynamic method}, we calculate the maximum H I Ly$\alpha$ intrinsic transit depth and He I triplet absorptions as shown in Figure \ref{figures/transmission_all_ism_2}, the values are listed in the label. As we can see, the H I Ly$\alpha$ is saturated because spanning a wide range of He/H ratios, depths are the same.

However, the optically thick ISM absorption will significantly affect the observed H I Ly$\alpha$ line. It makes the stellar radiation in the ISM H I Ly$\alpha$ absorption region almost undetectable, as shown in the spectra from MUSCLES Hubble Space Telescope Treasury Survey \citep{2016ApJ...824..101Y}. 
Since the ISM has an upwind Doppler shift of $\sim$26 km s$^{-1}$\citep{2013ApJ...779..130B}, consider the orbit motion of the observer on the Earth, and for a conservative estimation, we add an ISM H I Ly$\alpha$ absorption belt at the rest wavelength with the width of $\pm$ 30 km s$^{-1}$. Similarly, ISM deuterium (1215.3 $\rm \AA$) will also absorb the radiation, but its abundance is much poorer than the ISM hydrogen \citep{2005ApJS..159..118W}, usually optically thin, so we neglect its effect here. Consequently, with the ISM absorption, the maximum transit depths of H I Ly$\alpha$ are much weaker than the intrinsic transit depth as shown in the dashed lines in Figure \ref{figures/transmission_all_ism_2}.

In fact, the real impact parameter $b=0.15^{+0.12}_{-0.10}$ from \cite{2021MNRAS.507.2782O} is close to our assumption. And the simulation result in \cite{2020A&A...640A.134A} shows that in this deviation level, the impact parameter will not change the transit depth much, so the zero impact parameter assumption is acceptable.

Thus under He/H-dominated assumption, the escaping gas could be detected in transit spectroscopy, and the gas escaping predictions could be tested by observation. It's worth noticing that the atmosphere composition may not be He/H-dominated, for instance, \cite{2020ApJ...888L..21G} report non-detection of the $\pi$ Men c H I Ly$\alpha$ line in transmission spectroscopy with the Hubble Space Telescope’s Space Telescope Imaging Spectrograph. In that case, heavier elements (He, C, and O, etc.) observations are also valuable \citep{2020A&A...639A.109S}.

\section{Evolution history from NGPPS}\label{Evolution history from NGPPS}
\subsection{NGPPS model}
Planetary population synthesis is a powerful method, it can help us to reconstruct and understand the whole picture of planet formation and evolution. In the meantime, some parameters are unlikely to be measured in real observation, but it's possible to be retrieved from simulation, such as the planet's core and gas mass and radius. 

In the past decades, planetary population synthesis code (New Generation Planetary Population Synthesis, NGPPS) with the Bern model have been developed \citep{2021A&A...656A..69E, 2021A&A...656A..70E}. It begins the planets' formation and evolution from embryos within a 1D protoplanetary disk, considers the N-body interaction \citep{1999MNRAS.304..793C}, planet-disk co-evolution \citep{1973A&A....24..337S,2003ApJ...582..893M,2013A&A...549A..44F}, planet interior structure \citep{1986Icar...67..391B,1995ApJS...99..713S,2007ApJ...669.1279S}, and atmospheric escape \citep{2014ApJ...795...65J,2018AJ....155..214T} which in turn depends on the orbital location which decays over similar timescales due to stellar tides \citep{2011A&A...528A...2B}. Using the NGPPS data from \cite{2021A&A...656A..72B}, we obtain the formation and evolution time slices for 1000 planetary systems, with the 5 Gyr time coverage and host star mass of 0.7 $\rm M_{\sun}$.

\subsection{Similar NGPPS planets selection}

To find out the most similar planets in the synthetic data (0.7 $\rm M_{\sun}, R_{star}, P_{star}, log10(M_{p}), R_{p}, log10(a_{p})$), we define the $square\ error$ function ($se$) to describe the difference between the TOI-431 planets and NGPPS planets with the following form:
\begin{eqnarray}
    se = \Sigma_n\ \rm (para_{n\ TOI-431\ planet}-para_{n\ NGPPS})^2
\end{eqnarray}

Before calculating the $se$, it's important to balance the weights for all the parameters, to avoid the selection criteria tending to the large-number parameter, such as planet mass, and orbit distance, we normalize all the parameters with the $MaxMinScaler$ function. Assume we have a set $S_{old}$ of values $V_{old}$ of a parameter, the $MaxMinScaler$ function will calculate the new value $V_{new}$ of each element in the new set $S_{new}$, as: 

\begin{eqnarray}
    V_{new} = \frac{V_{old}-min(S_{old})}{max(S_{old})-min(S_{old})}
\end{eqnarray}
Then, we use $S_{new}$ to calculate $se$s, once the calculations have been completed, we select the top 100 minimum targets as the most similar NGPPG planets.

\subsection{Planet interior structure and gas envelopes lifetime}
\begin{figure}

	\includegraphics[width=\columnwidth]{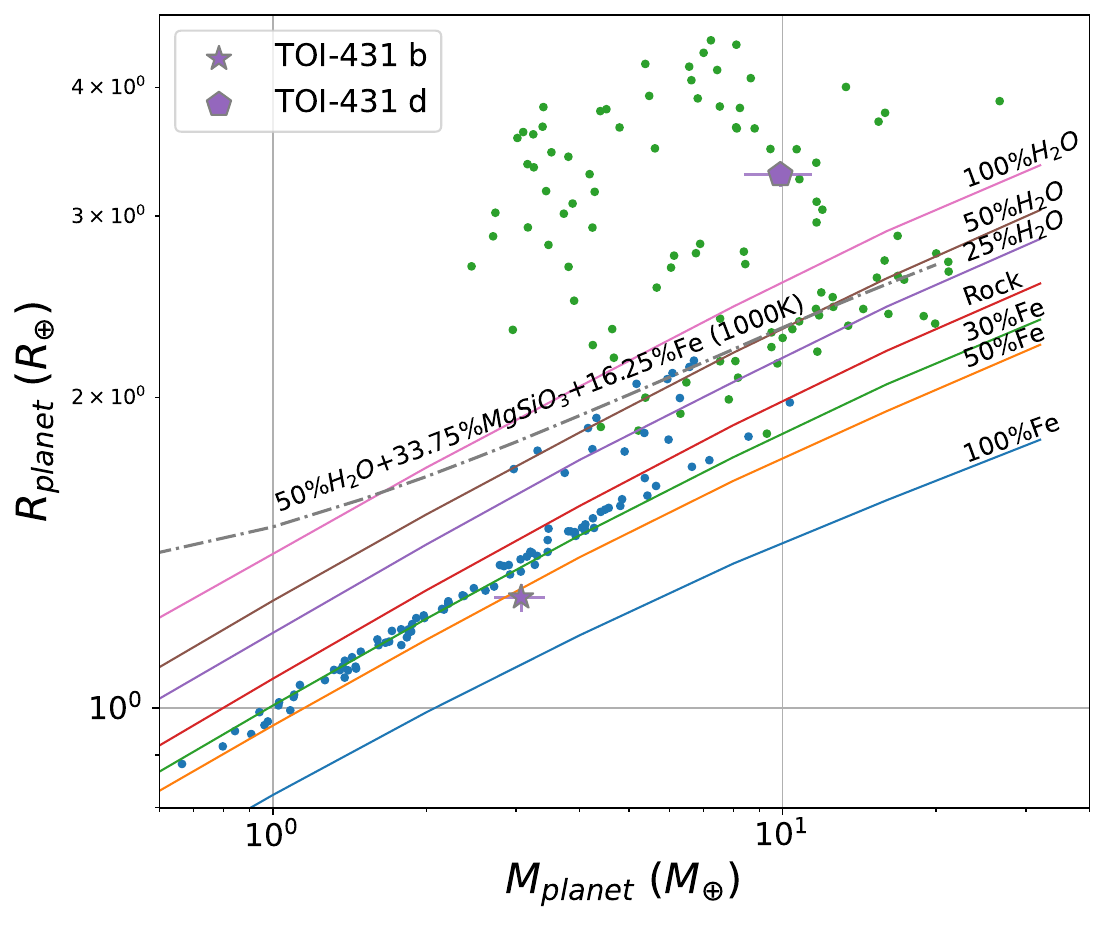}
    \caption{Mass-radius plot of TOI-431 planets and representative NGPPS planets at 5 Gyr with tow-layers mass-radius relations from \cite{2016ApJ...819..127Z, 2020A&A...643A.105H}. The blue dots represent the TOI-431 b-type planets and the green dots stand for the TOI-431 d-type planets in NGPPS.}
    \label{fig:m-r}
\end{figure}

In Figure \ref{fig:m-r}, we check the mass-radius relations for both TOI-431 b and d and their representative planets in NGPPS with some tow-layers mass-radius relations from \cite{2016ApJ...819..127Z, 2020A&A...643A.105H}. 
Obviously, almost all the TOI-431 b-like planets distribute along the 30\% Fe line, and below the rock line, which indicates that these planets should host an iron core inside. As to TOI-431 b itself, it is below the 50\% Fe line slightly, and denser than the representative planets, but not a pure iron core.
In the meantime, the TOI-431 d-like planets are much more diffuse, and the majority of them are larger than 100\% $\rm H_2O$ planets, for this reason, while they could still have some water, they must contain H/He gas envelopes.

\begin{figure}

	\includegraphics[width=\columnwidth]{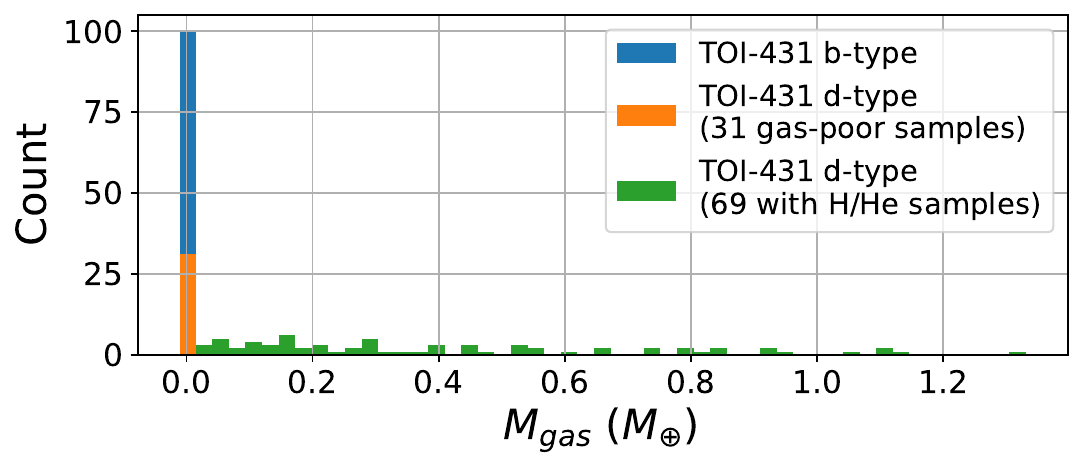}
    \caption{Gas mass distribution of representative NGPPS planets at 2 Gyr (obtained from the average result of 1 Gyr and 3 Gyr).}
    \label{fig:gas mass}
\end{figure}

Profit from the synthetic method, the interior structures are known, and we obtain the gas mass for the representative NGPPS planets as shown in Figure \ref{fig:gas mass}. It shows clearly that TOI-431 b-like planets are entirely gas-poor, TOI-431 d-like planets are likely to host a gas envelope with the average gas mass of 0.411 $\rm M_{\earth}$, though 31\% of them are nake cores.

By combining this average gas mass (gas-poor samples have been removed) with the TOI-431 d's $ATES$ mass-loss rate of $9.90^{+0.06}_{-0.02} \rm g/s$, the average lifetime of the gas envelope is $9.80^{+0.46}_{-1.26}$ Gyr, so it has a long enough lifetime for their gas envelope.
Moreover, 25, 50, 75, 100 percentiles TOI-431 d-like planets' gas mass exceeds 0.15, 0.30, 0.61, 1.33 $\rm M_{\earth}$, their corresponding lifetimes are $3.60^{+0.17}_{-0.47}$, $6.92^{+0.33}_{-0.89}$, $14.48^{+0.68}_{-1.87}$, $31.72^{+1.50}_{-4.09}$ Gyr, thus we can conclude that majority of them could maintain to the end of the stellar system, some of them even exceed the cosmos age. Considering the XUV radiation will be fainter and the $R_{XUV}$ will be smaller during the evolution, the real lifetime can be longer than our estimation, thus the gas envelope could be maintained until the host star dies.

On the contrary, if gas-rich TOI-431 d-like planets lose mass as TOI-431 b, the average gas lifetime is only $2.43_{-0.36}^{+0.1}$ Gyr, much shorter than the lifetime of a typical K-type star. For this reason, in theory, gas-poor TOI-431 b-like planets are almost impossible to maintain their gas envelopes to the end of their life, and they very likely are gas-poor planets.

\subsection{Evolution tracks}
\begin{figure}[h]
	\includegraphics[width=\columnwidth]{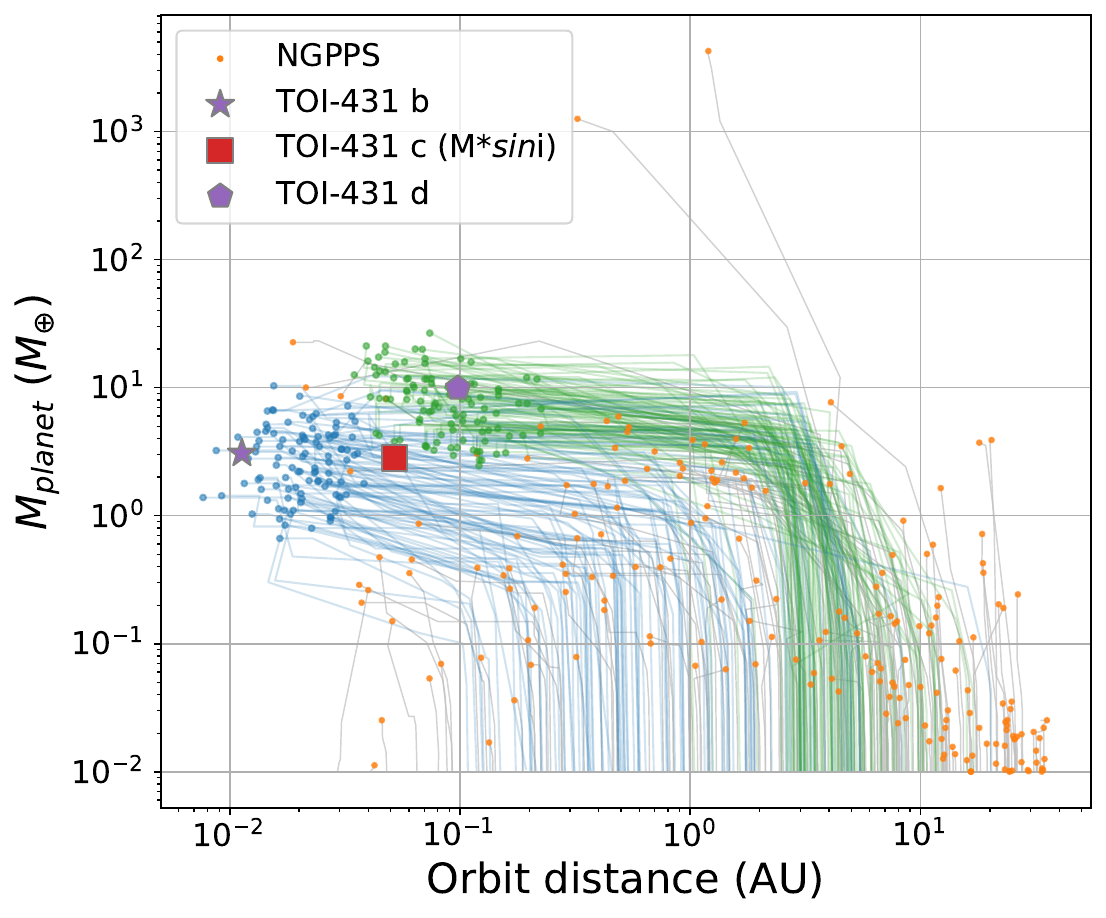}
    \caption{Evolution tracks (within 5 Gyr) of NGPPS planets and the similar NGPPS planets for both TOI-431 b and d, blue and green dots respectively.}
    \label{fig:ngpps tracks}
\end{figure}

\begin{figure}[h]
	\includegraphics[width=\columnwidth]{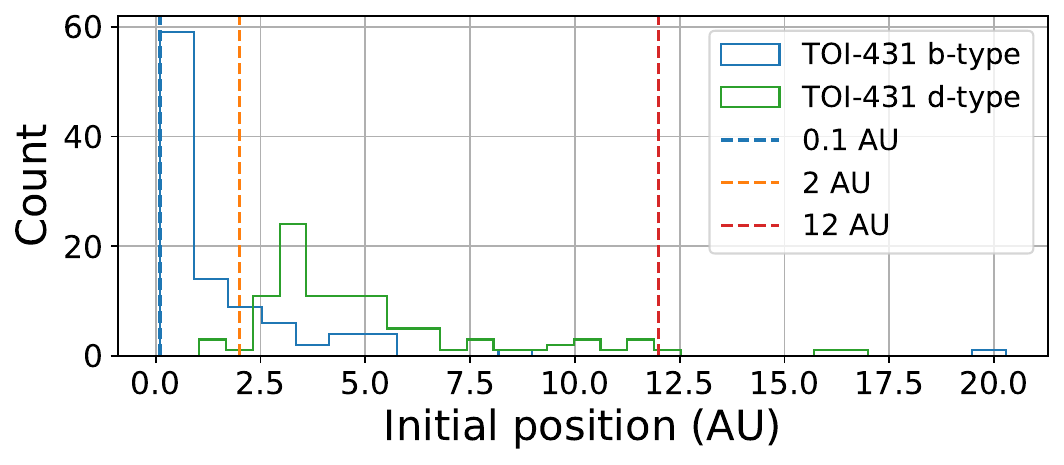}
    \caption{Histogram of the initial position of TOI-431 b and d-like planets in NGPPS, blue and green step respectively.}
    \label{fig: initial position}
\end{figure}

\begin{figure*}[t]
\centering
	\includegraphics[width=1\textwidth]{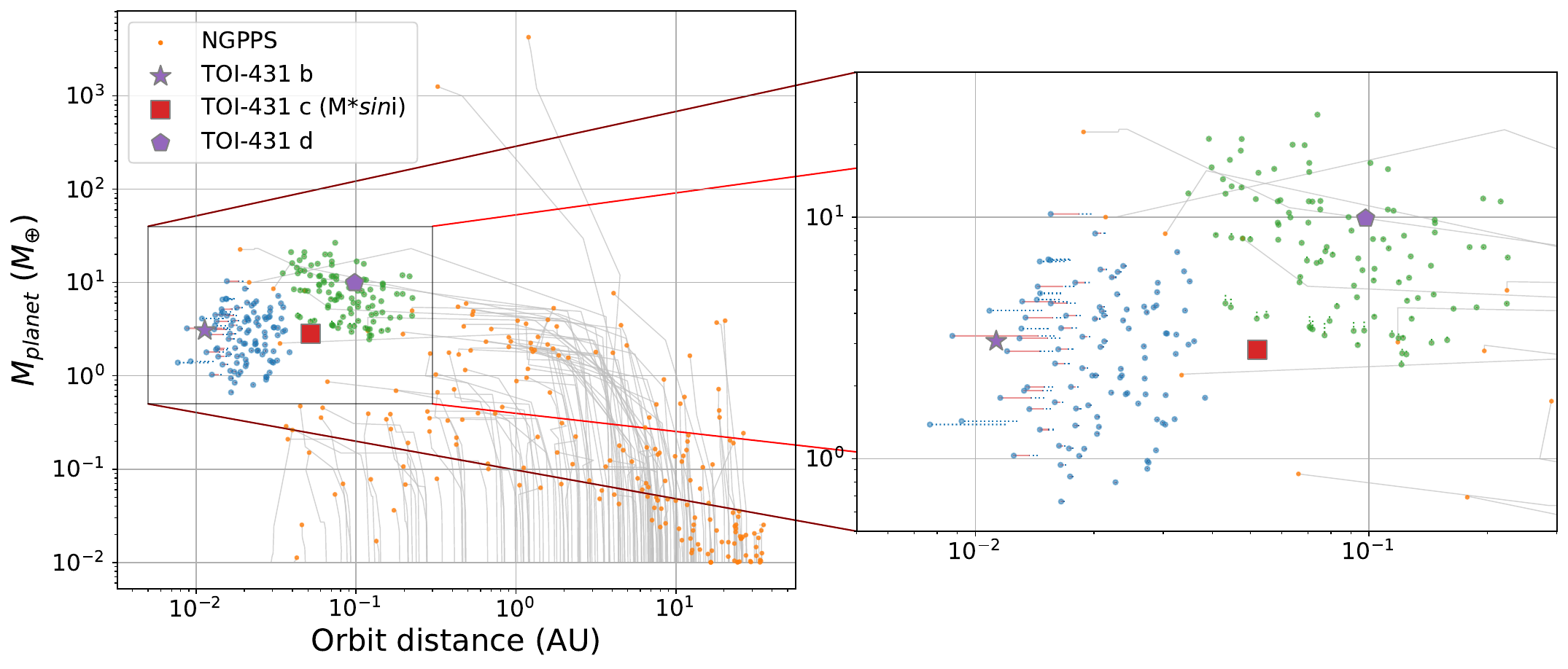}
    \caption{The tails show the late evolution tracks of similar planets, the dashed blue and green lines represent the age between 1Gyr to 3Gyr, and the solid red tails are for 3Gyr to 5Gyr. These tails show the system ages of 1.9Gyr and 5Gyr won't affect the results significantly, especially for TOI-431 d-like planets, the solid red tails are almost invisible.}
    \label{fig:7Msolar_a_M_sample100_end_zoomed}
\end{figure*}

The synthetic method can also help us study planets' history, as the evolution tracks in Figure  \ref{fig:ngpps tracks}. Interestingly, the histogram figure \ref{fig: initial position} clearly shows that TOI-431 b and d are formed in different regions in the system. TOI-431 b represents the relatively low-mass but hotter planets, they form within the range of 0.1-2 AU. For comparison,  TOI-431 d-like massive hot planets, formed in the outer region than TOI-431 b, between 2 and 12 AU. This phenomenon verifies the initial location can determine the fate of planets once more \citep{2023EPJP..138..181E}. And indicates the Fulton gap \citep{2017AJ....154..109F} origins from the gas-escaping in the phenomenon, and the deeper reason should be the XUV radiation and initial locations of both sides of planets, namely the mass distribution of protoplanetary disk. The planets formed in the inner region can migrate inward to more close-in orbits. Then they suffer violent XUV radiation for evaporation, and their mass-loss rates can reach $10^{10.51-9.14}\approx23.44$ times of TOI-431 d-like massive hot planets. Thus, they can hardly bind the gas envelopes and become lower-mass gas-poor planets. With the help of evolution tracks, we will also discuss the age issue in the following section.

\subsection{Model issue}\label{System age issue}

To match the correct sample, we should choose the corresponding time slice of NGPPS data. However, the system age is still unclear, it may induce some age-dependent bias.
Since the time slices of NGPPS data are 1Gyr, 3Gyr, and 5Gyr, and the estimated system age is 1.9Gyr or 5.1Gyr, the 1.9Gyr is basically in the middle of 1Gyr and 3Gyr. 
Considering that the planets will become more and more stable with the evolution of time, the changes per unit time will become smaller, as shown in Figure \ref{fig:7Msolar_a_M_sample100_end_zoomed}, the tracks within 3Gyr to 5Gyr are much shorter than the tracks within 1Gyr to 3Gyr. 
Additionally, in NGPPS, the evolution tracks between 1Gyr, 3Gyr, and 5Gyr are much shorter than the scales of similar planet clusters as shown in Figure \ref{fig:7Msolar_a_M_sample100_end_zoomed}, thus the bias of time slices choice in this range is not very significant. 
Furthermore, 3Gyr is closer to another predicted age of 5.1Gyr, to reconcile and account for the two possible ages, we match similar planets in the 3Gyr slice.

However, the NGPPS data isn't designed to directly retrieve the formation and evolution track of a certain planet. Also, the simulation has not yet been constrained by the observed XUV flux, and the simulated stellar mass is slightly lower than TOI-431, which may induce potential systemic bias. Still, with the help of the NGPPS data, we can explore wider parameter space, to have a larger picture of planet formation, especially, NGPPS including planet-disk co-evolution, N-body interaction, and planets migration. For comparison, \cite{2024MNRAS.530.3500K} use the observed X-ray flux as an anchor and single planet's parameters at a certain orbit to conduct diverse simulations, and discuss gas envelopes evolution for both TOI-431 b and d. Furthermore, our conclusions that TOI-431 b is naked, and TOI-431 d can hold its gas envelope for a long time are consistent with theirs well, and we both favor XUV-driven escaping mechanism.

\section{Discussion}
\subsection{He/H ratios}\label{He/H ratio}

\begin{figure}[h]
	\includegraphics[width=\columnwidth]{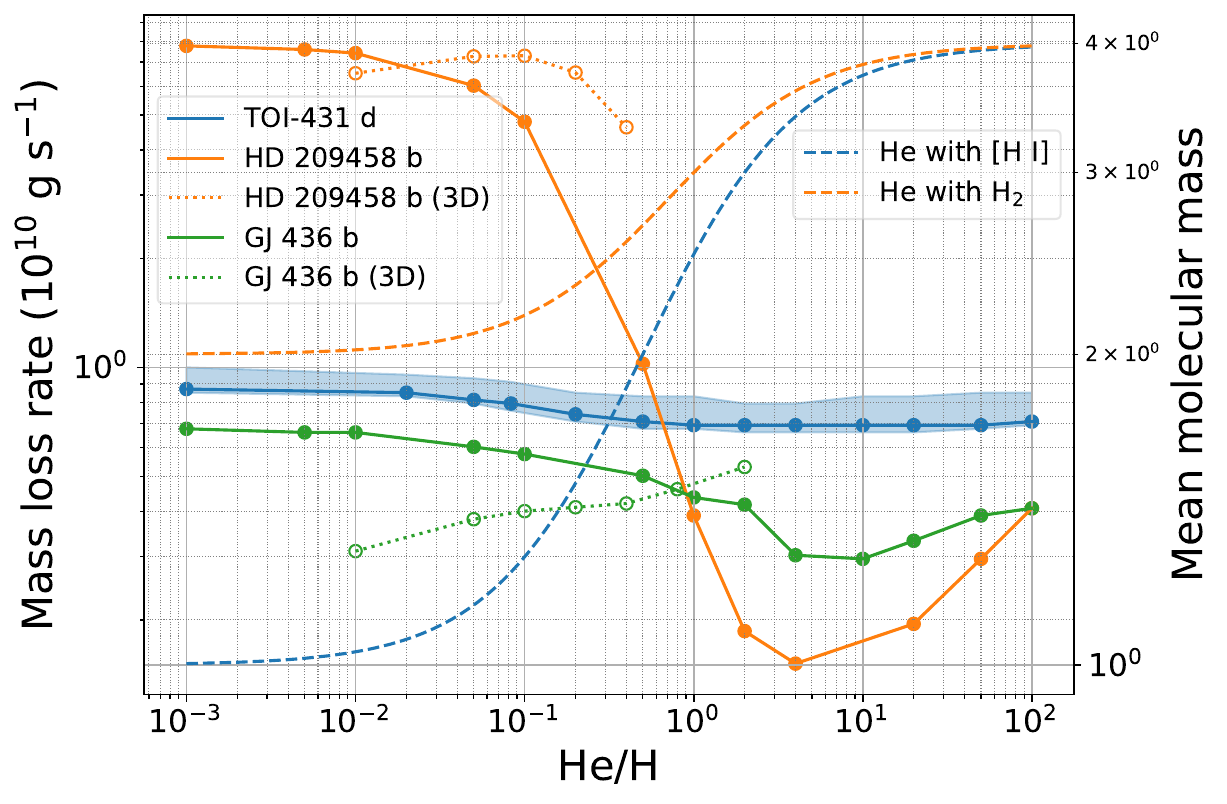}
    \caption{\textbf{Left axis:} relation of He/H ratio and mass loss rate of TOI-431 d, HD205498 b, and GJ436 b from $ATES$ (solid lines) and 3D model in \cite{2018MNRAS.481.5315S} (dotted lines); \textbf{Right axis:} relation of He/H ratio and average molecular mass of the atmosphere (dashed line).}
    \label{fig:heh_mass_dot_all}
\end{figure}

\begin{figure}[h]
	\includegraphics[width=\columnwidth]{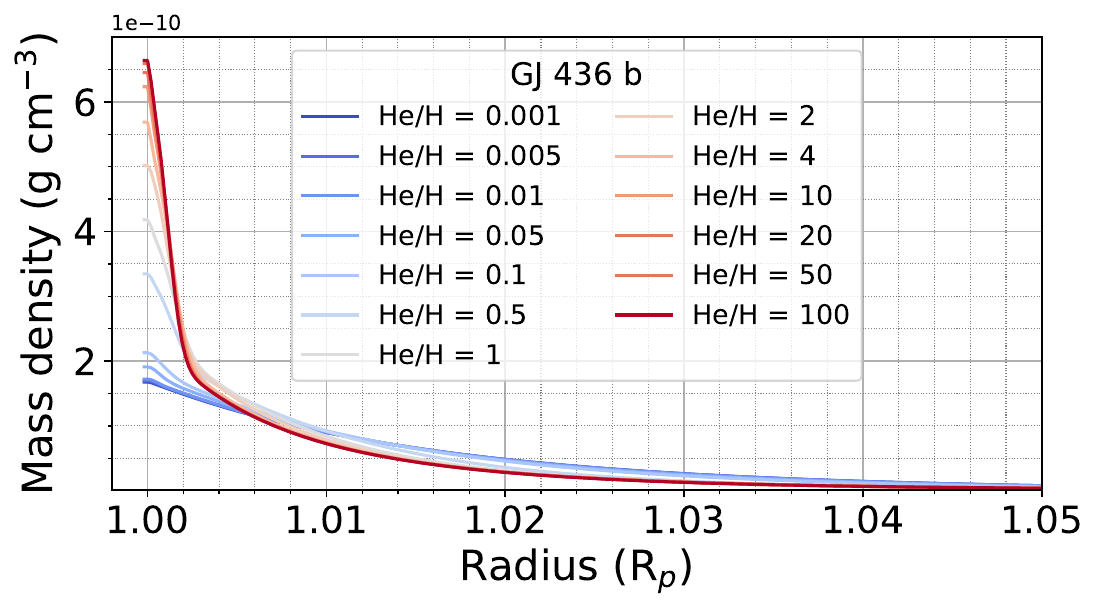}
    \caption{Density profiles of different He/H ratios for GJ~436 b from $ATES$. As the He/H ratio increases, the atmosphere will fall down quickly.}
    \label{fig:density_profiles}
\end{figure}

\begin{figure*}[t]
\centering
	\includegraphics[width=1\textwidth]{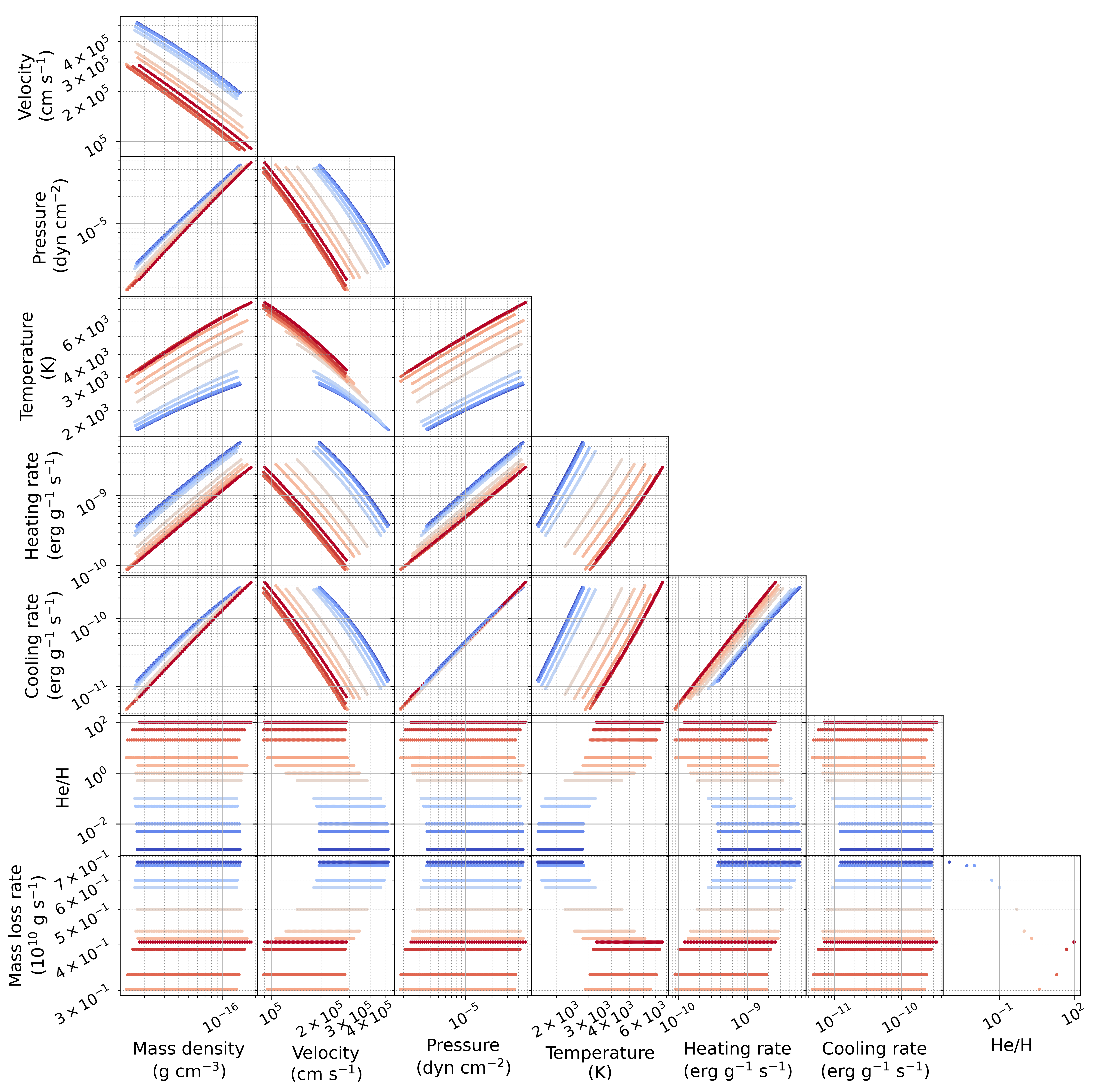}
    \caption{$ATES$ results of GJ~436 b with a set of He/H ratio. Each line shows the model parameters from 2.5 to 4.5 R${_{p}}$, the color represents the He/H ratio, and the corresponding values can be found in the He/H row.}
    \label{figures/GJ436b_corner_log_heh_color}
\end{figure*}

In principle, the gas escaping rate depends on the He/H ratio, H is lighter than He, so it can escape more easily. Thus, for a fixed temperature, decreasing the He/H ratio will result in an increasing mass-loss rate  \citep{2020A&A...636A..13L, 2021A&A...647A.129L}. 
However, it's unlikely to give the real He/H ratio without follow-up observation, but the mass-loss rate relies on the He/H ratio assumption, so we have to assume the He/H ratio. For warm Neptune GJ 3470 b, \cite{2021A&A...647A.129L} adopts a much lower He/H ratio of $1.5/98.5=0.015$, but \cite{2020A&A...638A..61P} gives higher $10/90 = 0.11$. We assume a set of He/H ratios for TOI-431 b, and $ATES$ results show the mass loss rates will decrease when He/H increases initially, and then increase with He/H, as shown in Figure \ref{fig:heh_mass_dot_all}.

\cite{2018MNRAS.481.5315S} also shows the mass loss rates may both increase and decrease with He/H ratio in 3D modeling of close-in exoplanets, as shown by the dotted line in Figure \ref{fig:heh_mass_dot_all}. They explain their curves in two stages. 
Firstly, before He-rich (He/H=0.05$\sim$0.1), the mass loss rate will increase with the He/H ratio, due to the increase in mean gas density, though outflow velocity also decreases. 
Secondly, when He/H$\geq $0.1, the atmosphere will fall off fast when He abundance increases further. In detail, there are two situations in this stage, on the lighter hot Neptune GJ~436 b, the atmosphere falls off faster than HD~209458 b as He/H increases, but the temperature and outflow velocity increase significantly as shown in Figures 7 and 8 in \cite{2018MNRAS.481.5315S}, so it can overcompensate the atmosphere fall off, and result in mass loss rate increase continually; In contrast, for massive hot Jupiter HD~209458 b, as the atmosphere falls off with He/H increases, the rise in the temperature and outflow velocity is weaker, thus the total mass loss rate will decrease. 

For comparing the mass loss rate results from $ATES$, we also use $ATES$ to model the same targets HD~209458 b and GJ~436 b in \cite{2018MNRAS.481.5315S} under a wider range of He/H ratios assumptions. In Figure \ref{fig:heh_mass_dot_all}, when He/H$\leq$10, GJ~436 b's mass loss rate will decrease as He/H increases, and if He/H exceeds 10, the mass loss rate will turn to increase. To find out the reason, we show the model atmosphere parameters under different He/H ratios in Figure \ref{figures/GJ436b_corner_log_heh_color}, and to avoid accidental situations at a specific radius, we include the data points between 2.5 to 4.5 R${_{p}}$. 
According to the slope, the relation between mass loss rate and He/H ratio can also be separated into three stages in the $ATES$ model. 
When the atmosphere is dominated by H (He/H$\leq$0.1), the atmosphere falls down slowly as He/H increases as shown in Figure \ref{fig:density_profiles}, thus the mass loss rate gradually decreases.
As the mean molecular mass increases significantly in the intermediate stage (0.1$\leq$He/H$\leq$10) as shown in Figure \ref{fig:heh_mass_dot_all}, the atmosphere falls down quickly as shown in Figure \ref{fig:density_profiles}. The temperature increases as the He/H ratio increases, but the radiative heating rate decreases and the mean molecular mass is enhanced quicker, so the outflow velocity decelerates significantly, decreasing the mass loss rate. 
Once the atmosphere becomes He-rich (He/H$\geq$10), the mean molecular mass becomes stable as shown in Figure \ref{fig:heh_mass_dot_all}. The radiative heating and cooling rates will increase with the He/H ratio, since the heating rate is one order higher than the cooling rate, the total heating rate increases, making the temperature, pressure, and velocity increase, therefore the mass loss rate increases again. For these reasons, both $ATES$ and \cite{2018MNRAS.481.5315S} can reflect the mass loss rate decrease caused by the atmosphere fall down, and the mass loss rate increase caused by the outflow velocity increase.

We can see similar mechanisms related to the atmosphere fall down and outflow velocity in the $ATES$ model and the 3D model in \cite{2018MNRAS.481.5315S}, but their behavior has some differences. The main reason is the composition assumptions in these models are different, the $ATES$ model assumes a primordial atmosphere composed of atomic hydrogen and helium (neglecting all forms of molecular hydrogen), but \cite{2018MNRAS.481.5315S} induce H$_{2}$. As the He/H ratio increases, the mean molecular mass in atomic H atmosphere increases much faster than molecular H${_{2}}$ atmosphere as shown in Figure \ref{fig:heh_mass_dot_all}, thus the velocity slopes are different.

In summary, comparing the results of $ATES$ and \cite{2018MNRAS.481.5315S} for HD~209458 b and GJ~436 b, although the trends of He/H are different, the range of mass loss rates is similar within the He/H overlap range of 2 to 3 orders of magnitude, with a mass loss rate difference of about 2 to 3 times, so the results are acceptable and have little impact on our conclusions. The assumed He/H ratio of 0.083 in this work, will give the relatively conservable gas escaping rates and He I triple absorption.

\subsection{TOI-431 b may experiencing Jeans escape}\label{toi-431 b discussion}
We can't obtain a convergence solution for TOI-431 b in Section \ref{Hydrodynamic method}. The reason could be it's experiencing Jeans escape but not hydrodynamic escape.
The gravitational potential is defined as \citep{2007A&A...472..329E}:
\begin{eqnarray}
    \Phi = &&-\frac{GM_{p}}{r}-\frac{GM_{*}}{a-r}\\ \nonumber
    &&-\frac{G(M_{p}+M_{*})}{2a^{3}}(r-a\frac{M_{*}}{M_{p}+M_{*}})^{2}
\end{eqnarray}
We find that TOI-431 b's gravitational potential is $10^{14.963}\ \rm erg/g$ and considering its XUV flux, it is located in the potential Jeans escape region in Figure 11 of \cite{2021A&A...655A..30C}. In this region, $ATES$ will fail to converge, because the atmosphere is likely undergoing Jeans escape rather than hydro-dynamical escape. Meanwhile, the energy limit method just gives an upper limit of the mass-loss rate, and TOI-431 b has very likely lost its gas envelope already, its real mass-loss rate is very uncertain.

\subsection{Stellar wind}

It's necessary to check whether the planets' escapes are confined by the stellar wind. In general, a planet's outflow will get in touch with the host star's outflows when escaping. Then stellar outflow puts ram pressure on the planet's outflow, it can shape or confine the planet's outflow. \cite{2009ApJ...693...23M, 2011ApJ...730...27A} have shown that if the stellar wind is strong enough, the escape rate can be reduced or prevented, and the planet can even obtain mass from stellar. The equilibrium region between the stellar wind and the evaporating planetary atmosphere is determined by the balance of total pressures, including both the ram pressure and outflow pressure, from the two winds. The key factor to consider is the location of the sonic point of the planetary outflow, which may occur either inside or outside of this stagnation region depending on the characteristics of the expanding planetary wind \citep{2020MNRAS.494.2417V}.

We assume an unmagnetized star and planets, to have a quick stellar wind confining check.
The $ATES$ code has given out the radial profiles of outflow pressure, temperature, and velocity for TOI-431 d, here, we need to find out the stellar wind ram pressure on planet orbit only. The ram pressure can be calculated by \citep{2020MNRAS.494.2417V}:
\begin{eqnarray}
    P_{ram,sw}(a_{orb})&=&\rho_{sw}(a_{orb})[u_{sw}^{2}(a_{orb})+\frac{GM_{*}}{a_{orb}}]\label{P_ram}
\end{eqnarray}
Thus the key to figuring out the $P_{ram}$ in eq.\ref{P_ram} is to determine the stellar wind density $\rho_{sw}(a_{orb})$ and speed $u_{sw}(a_{orb})$ around the planet. 

We adopt Parker's solar wind model \citep{2004suin.book.....S} to calculate the stellar wind speed $u_{sw}(a_{orb})$ as follows:

Estimate the coronal temperature $\bar{T}_{cor}$ from surface X-ray flux $F_{X}$ via the relation in \cite{2015A&A...578A.129J}:
\begin{eqnarray}
    \bar{T}_{cor}\approx0.11F_{X}^{0.26}
\end{eqnarray}
where $\bar{T}_{cor}$ in MK and $F_{X}$ in $\rm erg\ s^{-1}\ cm^{-2}$. Under the isothermal assumption, using the ideal gas equation of state and coronal temperature $\bar{T}_{cor}$ to calculate the isothermal sound speed $c_{s}$ in $\rm m\ s^{-1}$:
\begin{eqnarray}
    c_{s}=\sqrt{\frac{k\bar{T}_{cor}}{m_{H}}}=\sqrt{8.3*10^3*\bar{T}_{cor}}
\end{eqnarray}
where $k$ is the Boltzmann constant and $m_{H}$ is the mass of hydrogen atom. The critical radius $r_{c}$ in m where the stellar wind reaches sound speed is:
\begin{eqnarray}
    r_{c}=\frac{GM_{*}}{2c_{s}^{2}}
\end{eqnarray}
Take $c_{s}$, $r_{c}$ and planet's orbit distance $a_{orb}$ into the transcendental equation for the stellar wind speed $u_{sw}$ \citep{stellarwind}\footnote{\href{http://sun.stanford.edu/~sasha/PHYS780/SOLAR_PHYSICS/L22/Lecture_22_PHYS780.pdf}{Lecture 22 - Solar Wind}}:
\begin{eqnarray}
    (\frac{u_{sw}}{c_{s}})^{2}-2ln(\frac{u_{sw}}{c_{s}})=4ln(\frac{a_{orb}}{r_{c}})+4\frac{r_{c}}{a_{orb}}+3
\end{eqnarray}
and solve the equation by the graphical method, we can get the stellar wind speed $u_{sw}$ at planet orbit.
After that, the density of the wind $\rho_{sw}$ can be calculated via the mass conservation relation \citep{2020MNRAS.494.2417V}:
\begin{eqnarray}
    \dot{M}_{sw}=4\pi R^{2}\rho_{sw}u_{sw}
\end{eqnarray}
But, we need to estimate the stellar mass loss rate $\dot{M}_{sw}$ in addition. \cite{2018JPhCS1100a2028W} fit a mass loss rate and X-ray surface flux relation for GK dwarfs, giving the slope $a$ of $1.29 \pm 0.16$ in the form of:
\begin{eqnarray}
    lg(\frac{\dot{M}_{sw}}{R_{*}^{2}}) = a lg(F_{X})+b
\end{eqnarray}
where $\dot{M}_{sw}$ in $\dot{M}_{\sun}$, $R_{*}$ in $R_{\sun}$ unit, and $F_{X}$ in $\rm erg\ s^{-1}\ cm^{-2}$.
To find out the cutoff $b$, we refit the data in \cite{2018JPhCS1100a2028W} by the least squares method, and get the new formula which consists \cite{2018JPhCS1100a2028W}'s result well:
\begin{eqnarray}
    lg(\frac{\dot{M}_{sw}}{R_{*}^{2}}) = 1.3009 lg(F_{X})+5.8187
\end{eqnarray}
With this formula, we get TOI-431's mass loss rate of 6.32$\dot{M}_{\sun}$.
Finally, consider the $F_{X}$ uncertainty range, we get the stellar wind speed $u_{sw}$ of $\rm 380.212^{+29.725}_{-4.788}km\ s^{-1}$ and the $P_{ram,sw}$ of $\rm 1.177^{+0.734}_{-0.091}*10^{-5}dyn\ cm^{-2}$ for TOI-431 d.

\begin{figure}[h]
	\includegraphics[width=\columnwidth]{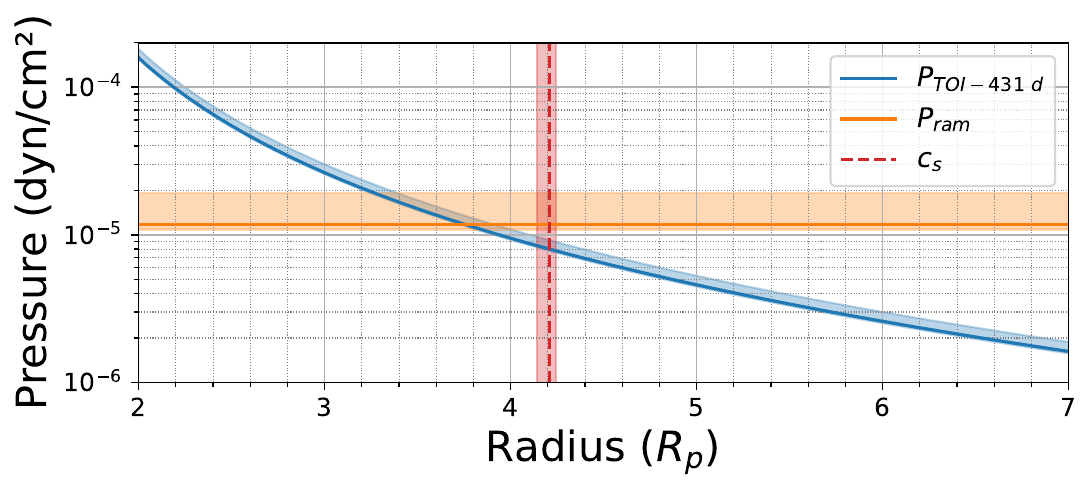}
    \caption{Compare the local stellar wind ram pressure (orange) and the TOI-431 d's outflow pressure (blue), their cross point is slightly inner than the sound speed layer (red) of planet outflow.}
    \label{fig:compare_stellar_wind_result}
\end{figure}

In Figure \ref{fig:compare_stellar_wind_result}, we compare the stellar wind ram pressure and TOI-431 d's outflow pressure, the cross point of these two pressures is slightly inner than the sound speed layer of the outflow, thus the stellar wind might confine TOI-431 d's mass loss rate, though it's very close to unconfined.

Moreover, the stellar wind is likely to decrease the transit depth and blow TOI-431 d's escaped gas to a further orbit, and form a comet-like gas tail, similar to the HD 209458 b \citep{2008A&A...483..933E}, HAT-P-32 b \citep{2023SciA....9F8736Z}, and Gliese 436 b \citep{2015Natur.522..459E}. 
The gas tail will give an asymmetric transit light curve. Because the gas in the tail has a lower orbit angular velocity, it will fall behind the planet. In this picture, the transit curve should be steep at the beginning side, and gentler at the end side.

\section{Conclusions}\label{Conclusions}

The XMM-Newton observation shows that the first two transit exoplanets in the TOI-431 system are suffering violent XUV radiation. 
Due to the extreme environment, the energy limit method gives the mass-loss rates for TOI-431 b and d of $10^{10.51^{+0.07}_{-0.02}}$ and $10^{9.14^{+0.07}_{-0.02}}$ g/s, the hydrodynamic code $ATES$ gives the mass-loss rates for TOI-431 d of $10^{9.84\sim 9.94}$ g/s for a set of He/H ratios, but TOI-431 b may experiencing Jeans escape.

Under the He/H-dominated assumption, TOI-431 d is expected the intrinsic H I Ly$\alpha$ absorption of $44.34^{+0.00}_{-0.00}\%$, and He I triplet depth up to $10.65^{+0.03}_{-0.00}\%$. However, the ISM absorption will reduce the H I Ly$\alpha$ depth down to 3-8\%, depending on the He/H assumed. Thus the gas escaping is possible to detect, but depends on the specific situation. Once gas escaping is confirmed in transit spectroscopy, the gas escaping as the reason for the Fulton gap can be supported. Also, the transit depth is likely reduced by stellar wind, if so, we can expect a decrease in the transit depth and an asymmetric transit light curve in observation, and then detect the gas tail.

With the help of the planetary population synthesis model NGPPS, we verify that TOI-431 b can not host a stable gas envelope, it should be a naked solid planet, and TOI-431 d has a much higher possibility of maintaining its gas envelope until the host star dies. 
From comparison to the formation and evolution tracks, we learn that TOI-431 b’s formation zone (0.1-2 AU) should lie closer to the star than that of TOI-431 d (2-12 AU). 
By analyzing tracks, we indicate that when explaining the Fulton gap as the result of the gas escaping, the intrinsic reason should be planets' birthplaces, which will determine how close these planets can migrate to the host star and then lose their mass, thereby form the statistic result as the Fulton gap. Thus, the two planets on opposite sides of the Fulton Valley are consistent with both formation or evolution leading to the observed feature (see also \cite{2020A&A...643L...1V}).
\\
\\
\textbf{Data Availability Statement:} Observation data used in this manuscript are publicly available in the XMM-Newton archive (Program PI: King, George, Observation ID: 0884680101). Other data are available upon request by contacting the corresponding author.
\\
\textbf{Acknowledgements:} This research has made use of the NASA Exoplanet Archive, which is operated by the California Institute of Technology, under contract with the National Aeronautics and Space Administration under the Exoplanet Exploration Program. Author JHJ acknowledges the support by the NASA exoplanet research program and by the Jet Propulsion Laboratory, California Institute of Technology, under contract by NASA. 


\bibliography{sample631}{}
\bibliographystyle{aasjournal}



\end{CJK*}
\end{document}